\documentclass[twocolumn,english]{revtex4-2}
\usepackage[T1]{fontenc}
\usepackage[latin9]{inputenc}
\usepackage{color}
\usepackage{amsmath}
\usepackage{graphicx}
\usepackage{comment}
\usepackage{graphicx}
\usepackage{sidecap}
\usepackage[colorlinks=true, allcolors=red]{hyperref}

\makeatletter
\usepackage{babel}

\makeatother

\usepackage{babel}
\begin{document}

\title{
{\color{black}Experimental observation of drift acoustic cnoidal waves in a magnetized plasma}}

\author{Tanmay Karmakar\textsuperscript{1,2}}
\email{tanmay.karmakar@ipr.res.in}

\author{Rosh Roy\textsuperscript{1,2}}

\author{Lavkesh Lachhvani\textsuperscript{1,2}}
\author{Raju Daniel\textsuperscript{1,2}}
\author{Bhoomi Khodiyar\textsuperscript{1,2}} 
\author{Prabal K. Chattopadhyay\textsuperscript{1,2}}
\author{Abhijit Sen\textsuperscript{1,2}}
\author{Sayak Bose\textsuperscript{3,a}}
\vspace{2cm}
\affiliation{\textsuperscript{1}Institute for Plasma Research, HBNI, Bhat, Gandhinagar, 382428, India} 

\affiliation{\textsuperscript{2}Homi Bhabha National Institute, Anushaktinagar, Mumbai, Maharastra 400094, India}

\affiliation{\textsuperscript{3}Columbia Astrophysics Laboratory, Columbia University, 550 West 120th Street, New York, New York 10027, USA}

\affiliation{\textsuperscript{a}Present address: Princeton Plasma Physics Laboratory, Princeton, New Jersey 08540, USA.}
%-----------------------------------------------------------
 
%-----------------------------------------------------------
\begin{abstract}  
\noindent
We report the experimental observation of highly nonlinear coherent structures in a linear magnetized plasma characterized by a strong background density gradient and significant $\mathbf{E}\!\times\!\mathbf{B}$ velocity shear 
under high ion-neutral collisionality. These structures, identified as drift acoustic waves, exhibit large normalized 
density fluctuations reaching amplitudes of up to $\sim 10\%$ and show periodic sawtooth -like waveforms. These observed waveforms are well described by cnoidal functions, corresponding to stationary nonlinear wave trains. Cnoidal waves are exact solutions of Korteweg-de Vries (KdV) - type equations, alongside the more commonly studied soliton solutions. To the best of our knowledge, this work presents the first controlled experimental observation of cnoidal wave trains in a highly collisional magnetized plasma through systematic variation of profile gradients.
These findings provide important new insights into the nonlinear evolution and saturation of drift acoustic waves in inhomogeneous, sheared, and collisional magnetized plasmas.

\end{abstract}
\maketitle

\section{Introduction}
\noindent  
The nonlinear saturated state of unstable drift waves can manifest either as incoherent turbulence \cite{horton1999drift,karmakar2026zonal, roy2025_experimental, burin_2005transition} or as coherent nonlinear structures such as vortices, solitons~\cite{saleem2024drift}, or periodic cnoidal-type waves \cite{kauschke1991nonlinear}. Many of these states are aptly described by solutions of a simplified two dimensional theoretical model equation -  the Hasegawa--Mima (HM)  equation~\cite{hasegawa1978pseudo}. The two dimensional coherent structures like drift vortices arise in a dispersive plasma media from a balance between nonlinear steepening and dispersive broadening of a nonlinear pulse. When the dynamics are predominantly one dimensional the HM equation reduces to the well known Korteweg-de Vries (KdV) equation. The KdV equation admits a rich variety of nonlinear solutions, including localized solitons and periodic steepened wave trains known as cnoidal waves. The exact form of cnoidal waves can be expressed in terms of Jacobi elliptic functions~\cite{Washimi1966,Sagdeev1966,InfeldRowlands2000}. In plasmas, such nonlinear periodic states can naturally emerge from the interplay of convective nonlinearity, plasma inertia, and dispersive effects \cite{hasegawa1978pseudo,Weiland2000}. In magnetized plasmas the underlying physical mechanism involves the nonlinear coupling of $\mathbf{E}\!\times\!\mathbf{B}$ drift and polarization drift: a perpendicular electric field produces transverse plasma motion, while temporal or spatial variations of the field introduce ion polarization currents that act as an effective inertia term \cite{hasegawa1978pseudo}. The competition between nonlinear advection and dispersive spreading therefore transforms initially sinusoidal perturbations into steepened periodic wave trains whose crests sharpen and troughs broaden, producing the characteristic cnoidal waveform \cite{InfeldRowlands2000,Weiland2000}. Cnoidal waves are
exact solutions of the Korteweg-de Vries (KdV) - type equations. They correspond to temporally periodic nonlinear states and are therefore particularly relevant to extended plasma systems where boundary conditions or continuous drive favor repeating structures rather than isolated pulses \cite{InfeldRowlands2000}. In bounded magnetized plasmas, velocity shear enhances the nonlinear convective term $u\,\partial u/\partial x$, causing faster regions of the wave to overtake slower ones and steepen the waveform; when dispersion remains finite the system evolves toward a periodic nonlinear equilibrium in the form of cnoidal oscillations \cite{Horton1999}. This balance can be characterized by a normalized shear parameter $S=\Delta V/C_s$, where $\Delta V$ is the velocity difference across the shear layer and $C_s$ the ion-acoustic speed; values $S\sim0.1$--$1$ favor nonlinear periodic states rather than linear sinusoidal waves or shock formation \cite{Horton1999}. Such conditions are often satisfied in laboratory magnetized plasmas and in tokamak edge and scrape-off-layer regions, where strong radial electric fields, density gradients, and $\mathbf{E}\!\times\!\mathbf{B}$ shear flows coexist \cite{Connor2004,Diamond2005} at high collisionality. Neutral particles can influence the turbulence and the formation of structures through collisional and other nonlinear processes \cite{Vranjes2010,Terasaka2022,Sessler1964,Burrell2005,Sasaki2020,Kosuga2020,Kobayashi2020,Kasuya2010}. In the edge and scrape-off layer of plasmas, they play an important role in modifying diverter heat load \cite{Myra2015,Krasheninnikov2017,Zholobenko2021}, affecting turbulence driven by plasma fueling \cite{Itoh2014,Ohshima2022}, and L to H transition by the gas puff \cite{Valovic2002,Singh2004}. Similarly, in space plasma environments such as the ionosphere and thermosphere, neutral interactions contribute to phenomena like super-rotation and anomolous resistivity \cite{Rishbeth1971,Papadopoulos1977}. Existence of cnoidal wave has been predicted for ion-acoustic and dust-acoustic modes, although experimental confirmation has remained limited~\cite{WashimiTaniuti1966,InfeldRowlands2000,KourakisShukla2005}. Recently, Liu \emph{et al.} provided direct and quantitative laboratory evidence of dust-acoustic cnoidal waves in an unmagnetized dusty plasma, demonstrating that measured nonlinear density oscillations evolve into steepened periodic profiles accurately described by Jacobi-elliptic KdV solutions, with the elliptic modulus increasing systematically with wave amplitude; this constitutes one of the clearest experimental validations of plasma cnoidal-wave theory in a non-magnetized plasma medium~\cite{liu2018experimental}. Kauschke \emph{et al.} analyzed nonlinear stationary periodic dispersive waves in magnetized plasma and derived harmonic coupling coefficients at $(2\omega,2k)$ and $(3\omega,3k)$, predicting cnoidal drift-waves ~\cite{kauschke1991nonlinear}.

\begin{figure*}[!hbt]
\includegraphics[width=\linewidth]{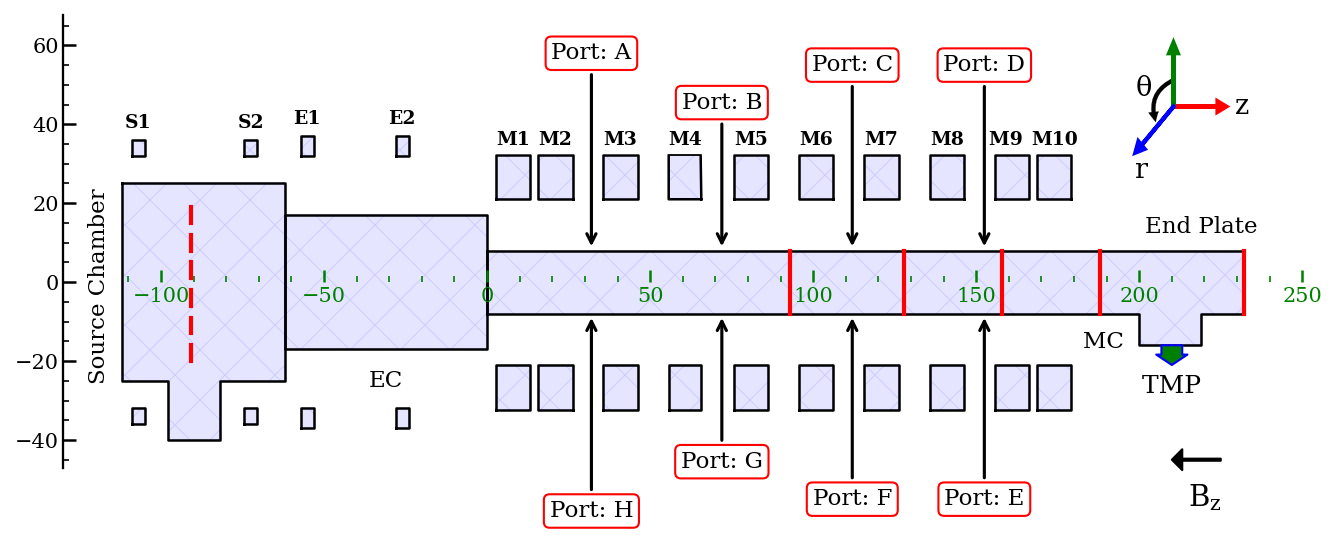} 
\caption{Filamentary discharge IMPED machine with radial diagnostic ports.
 }    
\label{IMPED schematics}  
\end{figure*}

\begin{figure*}[!hbt]
\centering
\includegraphics[width=0.9\linewidth]{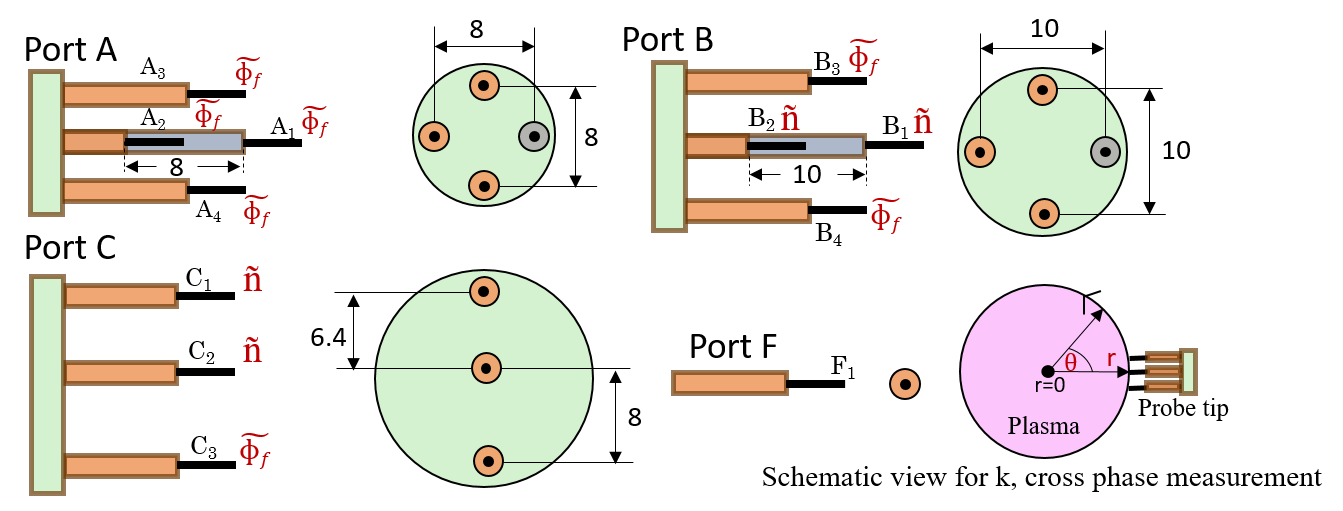} 
\caption{Probe configuration for measuring equilibrium profiles and fluctuation spectra. 
 }    
\label{Probe diagnostics}  
\end{figure*}

\noindent
Despite the extensive theoretical understanding of such nonlinear wave trains, controlled excitation of such nonlinear waves in magnetized plasmas has remained limited. In this work, we report the first experimental evidence of cnoidal-type wave trains in a linear magnetized plasma by systematic variation of the profile gradients. The observations are made under conditions of strongly enhanced normalized density and potential fluctuations localized within a narrow radial region. The density fluctuation signal exhibits a pronounced sawtooth-like waveform, characteristic of nonlinear steepening, while the corresponding potential fluctuations appear as localized, spike-like structures. The power spectrum of density fluctuations reveals a rich harmonic content with multiple higher harmonics and a characteristic beat frequency of $\sim1.1~\mathrm{kHz}$, consistent with the Doppler-shifted ion-acoustic frequency in the background $\mathbf{E}\!\times\!\mathbf{B}$ flow. These features indicate that the observed nonlinear periodic structures arise from the combined effects of ion-neutral collisionality and velocity shear-driven nonlinear advection. Furthermore, it is observed that the fundamental frequency of the cnoidal waves increases with magnetic field, highlighting the influence of plasma parameters on the nonlinear dynamics. Moreover, those cnoidal wave trains are strongly dependent on the profile gradient. Overall, these results provide strong experimental evidence that shear-driven nonlinear steepening, in the presence of collisional effects, can lead to the formation of drift-acoustic cnoidal wave structures in magnetized plasmas.
The paper is organized as follows. In Sec.~II, the experimental setup, diagnostics, and data acquisition techniques are described. In Sec.~III, the results are presented: Sec.~IIIA discusses the spectral characteristics at $550~\mathrm{G}$, $R_m=50$, and $2 \times 10^{-3}~\mathrm{mbar}$, while Sec.~IIIB examines the influence of radial profile gradients on the formation of cnoidal wave trains. In Sec.~IV, a physical interpretation of the observed nonlinear structures is provided. Finally, Sec.~V summarizes the main findings and presents the conclusions.

\section{Experimental Setup, diagnostics and Data Acquisition }
\noindent
Experiments were carried out in the Inverse Mirror Plasma Experimental Device (IMPED), a linear magnetized plasma device that produces plasma using a two-dimensional array of ohmically heated tungsten filaments, as shown in Fig.~\ref{IMPED schematics}. A detailed description of the device and operating conditions have been reported previously in Refs.~\cite{bose_2015inverse,Bose_2015_RSI,roy2025_experimental,roy2025experimental}. The basic plasma parameters, namely density ($n$), electron temperature ($T_e$), and plasma potential ($\phi_p$), are measured using a single Langmuir probe (SLP) of 0.5~mm diameter and 4.2~mm length installed at port~F (Fig.~\ref{IMPED schematics}); the probe geometry is chosen to minimize gyro-orbit effects. The analysis procedure for obtaining $n$, $T_e$, and $\phi_p$ and their radial profiles follows the method described in our earlier work~\cite{roy2025_experimental,roy2025experimental, karmakar_tanmay_10.1088/1741-4326/ae5723}. In fluctuation measurements, ion saturation current fluctuations ($\tilde{I}_s$) are treated as density fluctuations ($\tilde{n}$), while temperature fluctuations ($\tilde{T}_e$), measured using a triple Langmuir probe, are found to be much smaller than floating-potential fluctuations ($\tilde{\phi}_f$) in IMPED, consistent with earlier observations in similar small-scale devices~\cite{roy2025_experimental,perks_2022impact,oldenburger_2012dynamics}. To investigate the spectral structure in $(k_r, k_\theta)$ space, several probe arrays (Fig.~\ref{Probe diagnostics}) are employed at different radial ports. At port~C, cross-phase analysis between density and potential fluctuations is performed using signals from probes c1 and c3. In addition, the poloidal wavenumber ($k_\theta$) is estimated from the phase difference between signals measured at two poloidally separated tips (c1 and c2), enabling identification of the mode structure and underlying instability characteristics~\cite{smith_2007fast}. Furthermore, at ports~A and~B, four-tip probe arrays are employed to simultaneously measure fluctuations in density ($\tilde{n}$) and floating potential ($\tilde{\phi}_f$). These measurements allow independent estimation of both poloidal and radial wavenumbers ($k_\theta, k_r$), which are subsequently cross-validated using different combinations of fluctuating quantities and other spatial probe configurations. All probes which are mounted on a digitally controlled drive capable of radial scan with 0.2~mm positioning accuracy using a PID-based DC servo system~\cite{Roy2023_FED}. Fluctuating signals are being digitized using a PXIe-based high-speed data acquisition system at 1~MS/s with 1~s records (1~M points), ensuring reliable fluctuation spectra and a low noise floor.

\begin{figure}[]
\includegraphics[width=\linewidth]{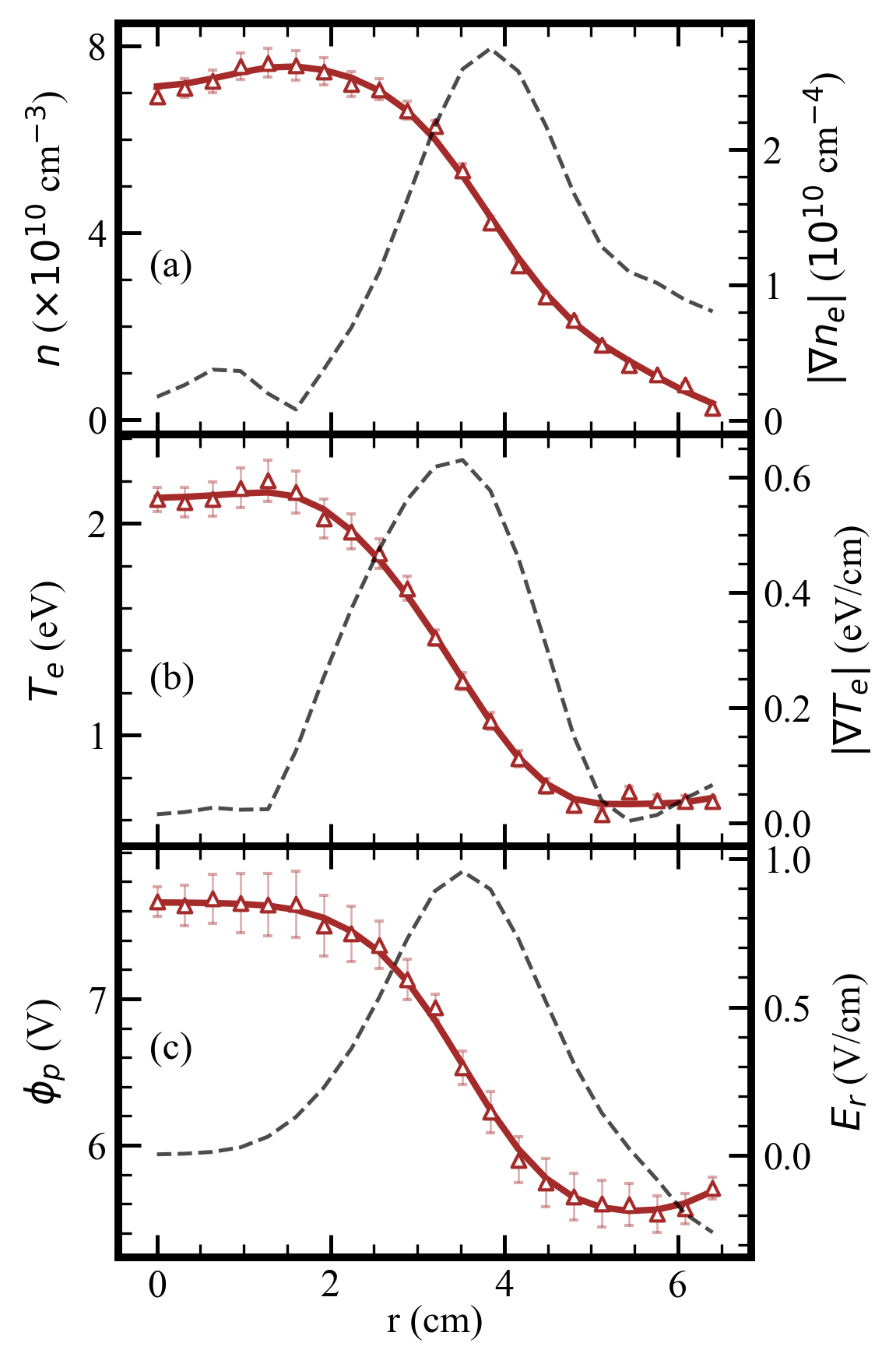} 
\caption{Radial mean profiles of (a) plasma density ($n$), (b) electron temperature ($T_e$), and (c) plasma potential ($\phi_p$) in solid lines and their respective gradients in dashed lines at $R_m=50,~550~G,~2\times10^{-3}~mbar$ pressure. 
}     
\label{fig:Mean profile at 550 G 2e-3 mbar}  
\end{figure}

\begin{figure}[]
\includegraphics[width=\linewidth]{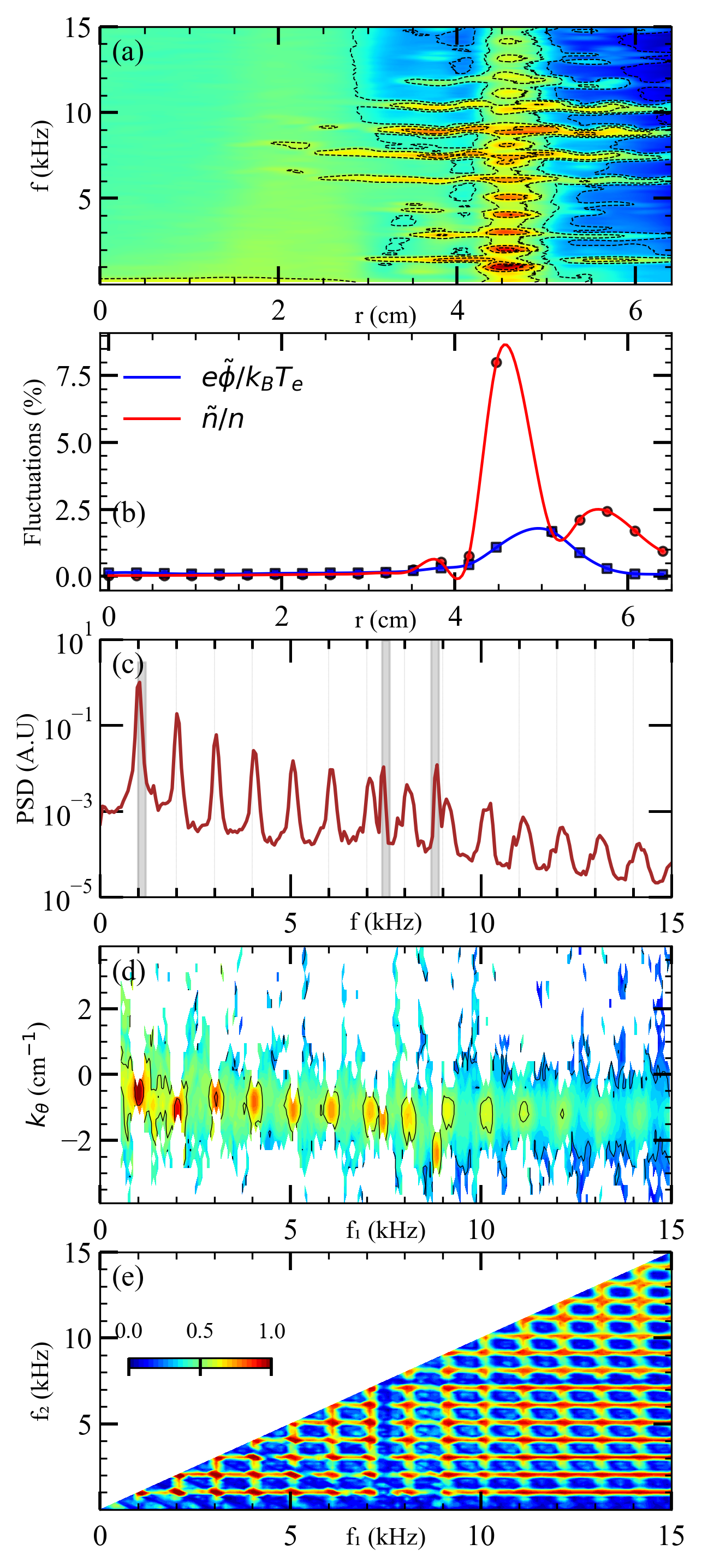} 
\caption{Spectral characteristics at $R_m = 51$, $B = 550$~G, and neutral pressure $2\times10^{-3}$~mbar. (a) Radial variation of the fluctuation spectra of density fluctuations ($\tilde{n}$). (b) Corresponding normalized potential and density fluctuation levels. (c) Auto power spectra of $\tilde{n}$ signal. (d) Poloidal wavenumber ($k_\theta$) spectra of $\tilde{n}$. (e) Auto-bicoherence of $\tilde{n}$ showing nonlinear mode coupling at 4.5 cm.}     
\label{fig: cnoidal wave at 550G}  
\end{figure}

\begin{figure}[]
\includegraphics[width=\linewidth]{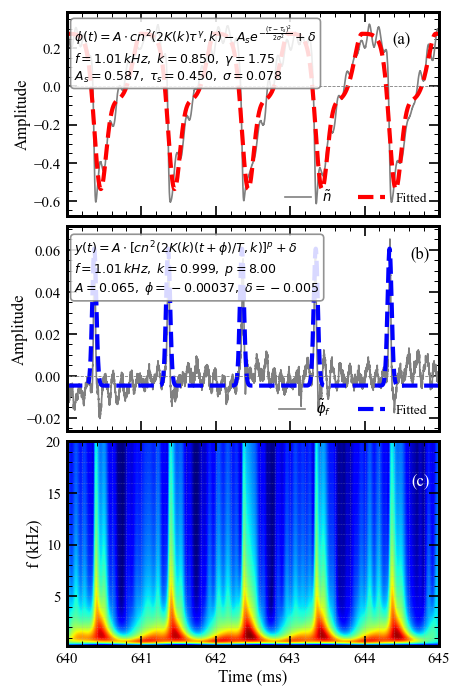} 
\caption{Comparison of experimental $\tilde{n}$ signal (gray) at 4.5 cm with two cnoidal wavefunctions. (a) with $\tilde{n}$ signal: $\phi = A\operatorname{cn}^2(2K\tau^\gamma) - A_s e^{-(\tau-\tau_s)^2/2\sigma^2} + \delta$. (b) with $\tilde{\phi}_f$ signal: $y = A[\operatorname{cn}^2(2K(t+\phi)/T)]^p + \delta$. Both fitted functions are consistent with a cnoidal-type solution. (c) Wavelet (Morlet) analysis of the $\tilde{n}$ signal.}    
\label{fig:comparison of fitting}  
\end{figure}
 
\section{Experimental results}
\subsection{Spectral characteristics at $R_m=50$, $B=550$~G, $P=2\times10^{-3}$~mbar}
\begin{figure}[]
\includegraphics[width=0.95\linewidth]{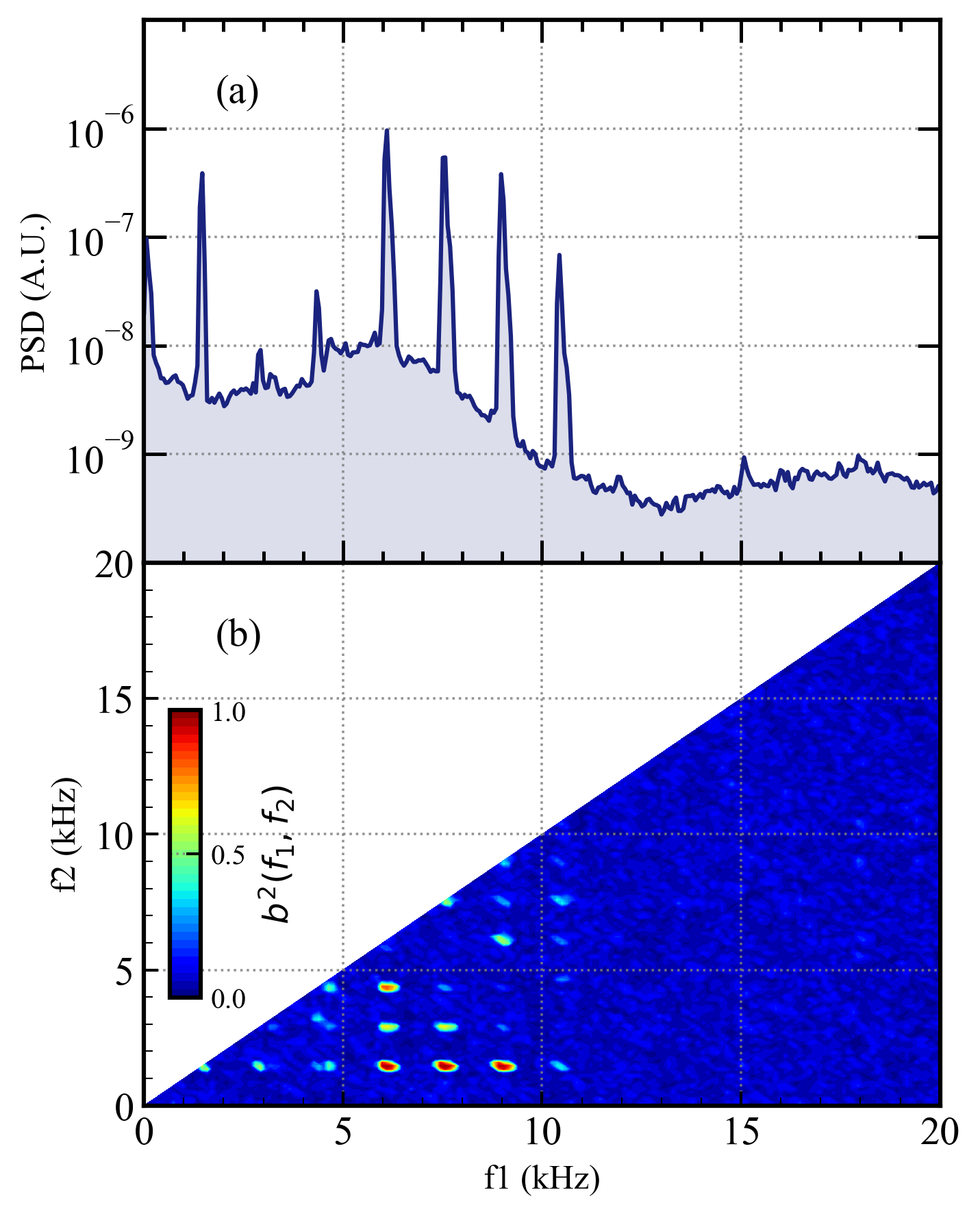} 
\caption{(a) Power spectral density of $\tilde{n}$ signal and (b) corresponding auto-bicoherence at $3.5 ~cm$ highlighting nonlinear wave coupling.}     
\label{fig: PSD_bicoherence at 550G at 3.2 cm}  
\end{figure}

\noindent
Radial mean profiles of plasma density ($n$), electron temperature ($T_e$), plasma potential ($\phi_p$) and their respective gradients at operating parameters of $R_m=50$, $B_m=550~\mathrm{G}$, and neutral pressure $1\times10^{-4}~\mathrm{mbar}$ are shown in Fig.~\ref{fig:Mean profile at 550 G 2e-3 mbar}. The peak density reaches $7\times10^{10}~\mathrm{cm^{-3}}$ with a maximum gradient of $3\times10^{10}~\mathrm{cm^{-3}}$, while the peak $T_e$ and $\phi_p$ are $2.2~\mathrm{eV}$ and $7.5~\mathrm{V}$, respectively. At this neutral pressure, ratio of ion neutral collision ($f_{in}$) to the ion gyro frequency ($f_{ci}$) is $\sim 1.12$ (assuming $T_i \sim 0.01~eV$). The locations of maximum gradients in density, temperature, and plasma potential occur at $r=3.8~\mathrm{cm}$ and $3.5~\mathrm{cm}$, respectively. To see the effect of profile gradients on the fluctuations, spectral analysis corresponding to these plasma profiles have been done (Fig.~\ref{fig: cnoidal wave at 550G}). The radial variation of fluctuation spectrum of $\tilde{n}$ is shown in Fig.~\ref{fig: cnoidal wave at 550G}a. It shows the presence of multiple coherent fluctuations at 4.5 cm, downstream of the velocity shear layer. The normalized fluctuation level $\tilde{n}/n$ (Fig.~\ref{fig: cnoidal wave at 550G}b) also shows a sudden jump at $r\approx4-5~\mathrm{cm}$, just downstream of the velocity shear and density gradient region, coinciding with the appearance of multiple coherent frequency modes. The auto power spectrum of $\tilde{n}$ signal shows spectra consists of multiple coherent frequencies (Fig.~\ref{fig: cnoidal wave at 550G}c). It consists of a fundamental mode of 1.1 kHz and another two distinct modes between 7-9 kHz frequency band. The $S(k,\omega)$ spectrum (Fig.~\ref{fig: cnoidal wave at 550G}d) also reveals pronounced power distributed across several neighboring poloidal modes extending up to $18~\mathrm{kHz}$, indicating the possible presence of nonlinear triadic interactions among these modes. This has been confirmed by the auto-bicoherence analysis(Fig.~\ref{fig: cnoidal wave at 550G}e) \cite{kim_2007_digital}. To identify these coherent fluctuations, it is found that the fundamental beat frequency is  $\approx1.1~\mathrm{kHz}$. This mode propagates in electron diamagnetic direction (clockwise relative to the plasma source). At this radial location, $T_e\approx1~\mathrm{eV}$ gives an ion acoustic speed of $\sim1500~\mathrm{m/s}$.  The $E \times B$ velocity in this radial range is $1636~\mathrm{m/s}$ and is directed anticlockwise in the same reference frame. Since the mode propagation is opposite to the $E \times B$ rotation, the Doppler-corrected frequency of the 1.1 kHz mode is estimated to be $\sim 950~\mathrm{Hz}$ (using $k_\theta = 0.44~\mathrm{cm^{-1}}$), which is close to the observed value of $\sim 1.1~\mathrm{kHz}$. This agreement indicates that the fundamental mode corresponds to an ion acoustic mode ($m=2$). This mode has axial wavenumber $(k_z)~\sim 0$. The $\tilde{n}$ signal exhibit a periodic sawtooth-like waveform with nonlinear steepening, whereas the $\tilde{\phi}_f$ signals appear as localized, spike-like patterns (Fig.\ref{fig:comparison of fitting}(a-b)). Wavelet analysis (Fig.\ref{fig:comparison of fitting}c) of the $\tilde{n}$ signal indicates that most of the fluctuation energy is confined within $\sim 5$ kHz band, indicating the dominance of large-scale low-frequency modes. The phase difference between $\tilde{n}$ and $\tilde{\phi}_f$ indicates a deviations from the linear Boltzmann relation, with density forming a continuous nonlinear wave train and potential responding in a more localized manner, consistent with a nonlinear drift wave regime. Such signals are a characteristic of nonlinear wave steepening, where the crest of the wave propagates faster than the trough due to asymmetric ion motion in the wave electric field. Such nonlinear evolution can arise from the combined effects of strong ion neutral collisions and significant velocity shear ($dV_E/dr\approx9\times10^{2}~\mathrm{s^{-1}}$), which together can excite nonlinear structure. The waveform is well reproduced by a cnoidal-type fit of the form  $\phi(t)=A\,\mathrm{cn}^{2}(2K(k)\nu t,k)$ with elliptic modulus $k=0.85$, indicating a nonlinear periodic structure approaching the soliton-train regime. To account for the pronounced localized density dips, an additional Gaussian correction term is included in the fit, which captures the asymmetric spike-like features present in the experimental signal. The corresponding potential fluctuation $\tilde{\phi}$ displays narrow periodic positive spikes and is also well fitted by a cnoidal function $y(t)=A\,[\mathrm{cn}^{2}(2K(k)(t+\phi)/T,k)]^{p}+\delta$, producing sharply peaked pulse-like structures consistent with the observed waveform. The elliptic modulus $k$ serves as a measure of nonlinearity, with values approaching unity corresponding to increasingly localized soliton-like structures; the obtained values ($k\approx0.85, 0.99$) therefore signify a nonlinear periodic wave with close resemblance to the cnoidal solutions of the Korteweg de Vries (KdV) equation. From Lakhin \textit{et al.}'s theory (1988) \cite{lakhin1988drift}, the time period of the cnoidal wave can be estimated as $T=\ln\!\left(\frac{16}{1-k}\right)\sqrt{\frac{12}{C_s^2\,(\partial \kappa_n/\partial x)\,A_{\mathrm{tot}}}}$, where $k$ is the elliptic modulus characterizing the waveform nonlinearity (here $k=0.85$), $A_{\mathrm{tot}}$ is the normalized fluctuation amplitude ($A_{\mathrm{tot}}=\tilde{n}/n=0.08$), $\partial \kappa_n/\partial x$ represents the spatial variation of the inverse density scale length (approximated as $\sim 1/L_n^2$ with $L_n=2.2~\mathrm{cm}$), and $C_s$ is the ion sound speed ($C_s=1500~\mathrm{m/s}$). Substituting these values, we obtain $T\approx0.84~\mathrm{ms}$, corresponding to a frequency of $\sim1.2~\mathrm{kHz}$. This theoretical estimate is in close agreement with the experimentally observed frequency of $\sim1.1~\mathrm{kHz}$, supporting the identification of the measured structures as cnoidal-type nonlinear drift waves. Within each oscillation of the $\tilde{n}$ signal, a higher-frequency fluctuation is also observed, exhibiting large density fluctuation in the $6$--$10~\mathrm{kHz}$ range (Fig. \ref{fig:comparison of fitting}a) and propagating in the electron diamagnetic direction. The measured poloidal wavenumber is $k_\theta = 1.5~\mathrm{cm^{-1}}$. Using this, the Doppler-corrected electron diamagnetic frequency is estimated as $f_{De} \approx 9.1~\mathrm{kHz}$, which is in close agreement with the experimentally observed frequency. The parallel wavenumber is found to be $k_z \sim 0.02$ around this frequency range. These observations confirm the presence of drift modes in the frequency range $6$ -$10~\mathrm{kHz}$~\cite{schroder2005drift}.Furthermore, the auto-bicoherence analysis (Fig.~\ref{fig: cnoidal wave at 550G}d) of the $\tilde{n}$ signal reveals multiple prominent nonlinear interactions, including higher harmonics of the $1.1~\mathrm{kHz}$ mode, as well as nonlinear coupling between the $\sim 1.1~\mathrm{kHz}$ ion-acoustic mode and the $6$--$10~\mathrm{kHz}$ drift-wave modes. To further substantiate this observation, auto-bicoherence analysis has been performed at a nearby radial location ($r = 3.5~\mathrm{cm}$). The nonlinear coupling between these modes is consistently observed at this location as well, as illustrated in Fig.~\ref{fig: PSD_bicoherence at 550G at 3.2 cm}(a--b). This confirms a nonlinear interaction between ion-acoustic and drift waves, resulting in the formation of a coupled nonlinear cnoidal structure. The density fluctuation level is significantly enhanced where this nonlinear structure is formed, with amplitudes rising up to $\sim 15$ times higher than those in the adjacent spatial regions. We refer to this emergent coherent state as a ``drift-acoustic cnoidal wave.'' Since the background profile gradient plays an important role in driving such nonlinear modes, it is worthwhile to investigate its influence in the present scenario. This will be discussed in the following section.  

\begin{figure*}[]
\includegraphics[width=1\linewidth]{revised_fig_09_10_25/DC_profiles_with_raw_data_650G_Rm_variation.png} 
\caption{Radial mean profiles of (a) density ($n$), (b) electron temperature ($T_e$), (c) plasma potential ($\phi_p$), (d) density gradient, (e) electron temperature gradient, and (f) radial electric field for $R_m = 50$ (red-solid), $33$ (green-dashed), and $17$ (blue-dotted), respectively. Panels (g)--(i) show the raw $\tilde{n}$ signals for $R_m = 50$, $33$, and $17$, respectively. Experiments are performed at $B = 650\,\mathrm{G}$ and neutral pressure $p = 2 \times 10^{-3}\,\mathrm{mbar}$.} %Colors correspond to $R_m = 50$ (red), $33$ (green), and $17$ (blue).}     
\label{fig:Mean profile at 650 G 2e-3 mbar}  
\end{figure*}

\begin{figure*}
\includegraphics[width=1\linewidth]{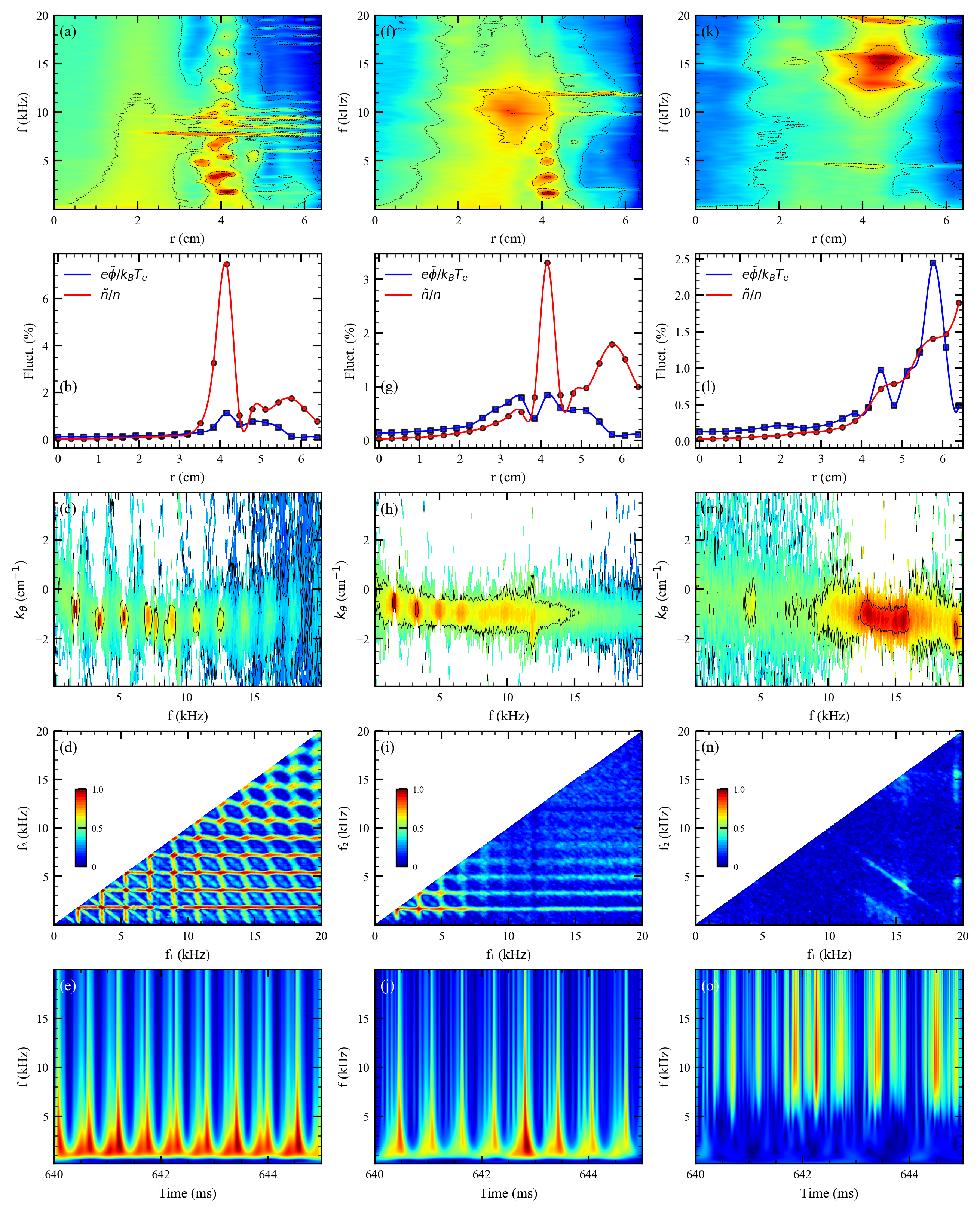} 
\caption{Effect of $R_m$ variation (50, 33, and 17) on plasma fluctuation characteristics at $650~G$. 
For $R_m = 50$ (a to e), $R_m = 33$ (f to j), and $R_m = 17$ (k to o): 
(a, f, k) radial power spectra of the density fluctuation ($\tilde{n}$); 
(b, g, l) normalized fluctuation levels of potential and density; 
(c, h, m) spatiotemporal spectra $S(k,\omega)$; 
(d, i, n) auto-bicoherence analysis; 
(e, j, o) wavelet analysis of the density fluctuation ($\tilde{n}$).}    
\label{fig:Rm variation of cnoidal wave at 650G 2e-3 mbar}  
\end{figure*}

\begin{figure}
\includegraphics[width=\linewidth]{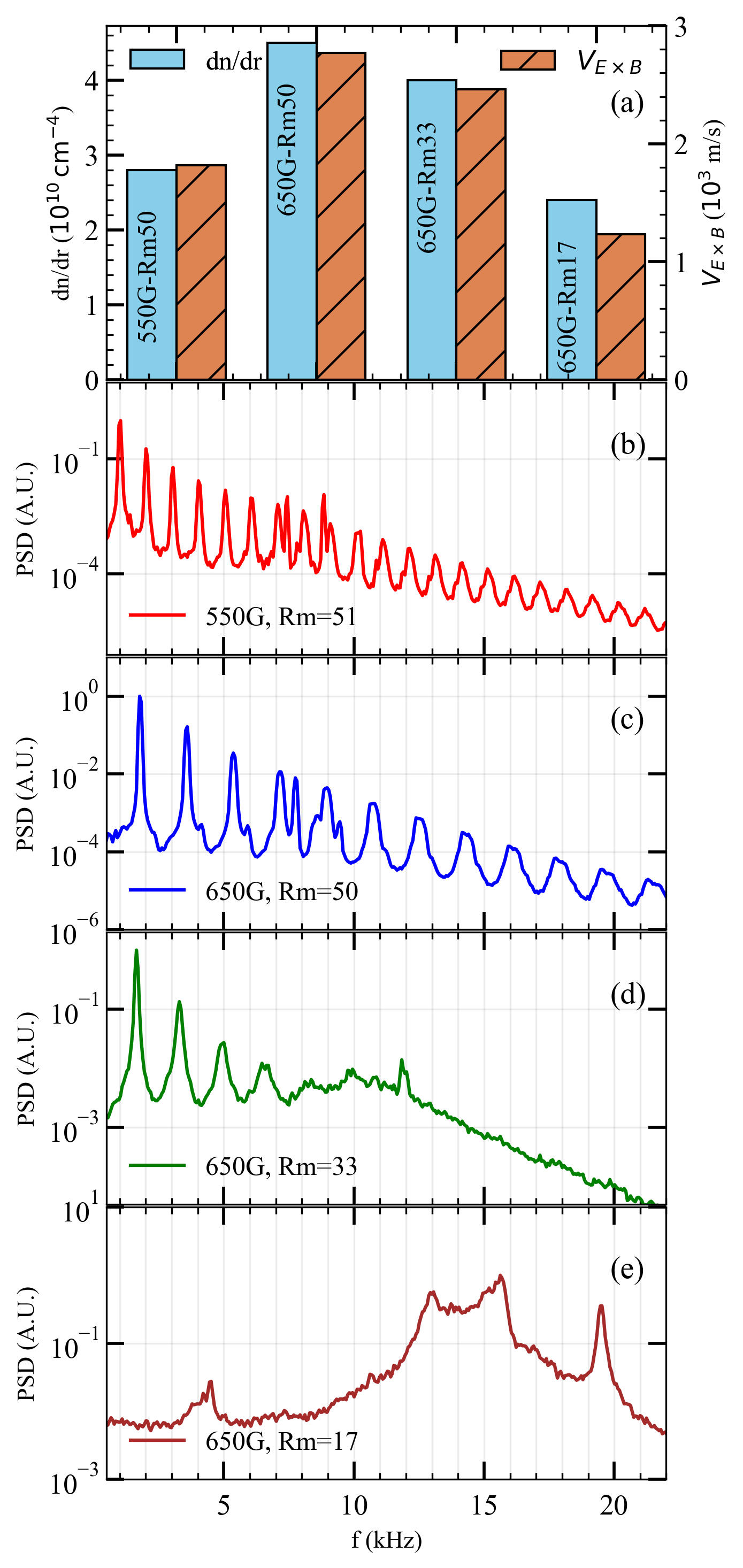} 
\caption{(a) Density gradient ($dn/dr$) and $E \times B$ velocity for different magnetic field strengths ($B_m$) and $R_m$. 
Evolution of the power spectral density (PSD) of normalized density fluctuations ($\tilde{n}$), illustrating the transition from coherent harmonic modes to broadband spectra: 
(b) $B_m = 550~\mathrm{G},~R_m = 51$, 
(c) $B_m = 650~\mathrm{G},~R_m = 50$, 
(d) $B_m = 650~\mathrm{G},~R_m = 33$, and 
(e) $B_m = 650~\mathrm{G},~R_m = 17$.}    
\label{fig: PSD and profile variation}  
\end{figure}

\subsection{Influence of Radial Profile Gradients}
\noindent
The influence of background profile gradients on the fluctuation dynamics is investigated by systematically varying the magnetic field and magnetic field ratio $R_m$ (ratio of main magnetic field to the source magnetic field), which effectively modifies the available free energy in the plasma. When the main magnetic  field is increased from $550~\mathrm{G}$ to $650~\mathrm{G}$ at $R_m=50$, the background plasma profiles become significantly steeper, as shown in Fig.~\ref{fig:Mean profile at 650 G 2e-3 mbar}(a--f). In particular, the plasma density increases from $2.6\times10^{10}~\mathrm{cm^{-3}}$ to  $4.5\times10^{10}~\mathrm{cm^{-3}}$, indicating an enhancement of the density gradient. At the same time, the maximum radial electric field increases from $\sim100~\mathrm{V/m}$ to $\sim180~\mathrm{V/m}$, which strengthens the $E\times B$ flow shear and increases the peak velocity shear from $1818~\mathrm{ms^{-1}}$ to $2769~\mathrm{ms^{-1}}$. These stronger gradients provide additional free energy for drift-wave instability, leading to a clear modification in the fluctuation spectra of $\tilde{n}$ (Fig.~\ref{fig: PSD and profile variation}b). In this regime, the fundamental fluctuation frequency shifts from $1.1~\mathrm{kHz}$ to $1.9~\mathrm{kHz}$ and a sudden increase in the normalized density fluctuation level is observed. Furthermore, the peak density gradient location shifts radially inward from $\sim3.8~\mathrm{cm}$ to $\sim3~\mathrm{cm}$, and a similar inward shift is seen for the location of maximum velocity shear. As a consequence, the spatial localization of the coherent modes also moves inward, with multiple coherent modes in the $\tilde{n}$ fluctuation spectra appearing around $\sim4.16~\mathrm{cm}$ instead of $\sim4.5~\mathrm{cm}$ in the $550~\mathrm{G}$, $R_m=50$ case. This observation highlights the close correlation between background gradients and the radial localization of the fluctuation spectra. The corresponding $k_\theta$ spectrum  (Fig.~\ref{fig:Rm variation of cnoidal wave at 650G 2e-3 mbar}c) and the bi-coherence analysis (Fig.~\ref{fig:Rm variation of cnoidal wave at 650G 2e-3 mbar}d) further reveal the presence of distinct coherent modes and their nonlinear interaction with harmonic components, indicating a strongly nonlinear wave - wave coupling process. The wavelet analysis (Fig.~\ref{fig:Rm variation of cnoidal wave at 650G 2e-3 mbar}e) shows that most of the fluctuation power remains concentrated within the $5$-$6~\mathrm{kHz}$ band, suggesting the persistence of coherent nonlinear structures in time. The effect of reducing the profile gradients is then examined by decreasing the $R_m$ from $R_m=50$ to $R_m=33$. This reduction weakens both the density gradient and the velocity shear, thereby reducing the available free energy that drives drift-wave instabilities (Fig. \ref{fig:Mean profile at 650 G 2e-3 mbar}(d,f)). As a consequence, the normalized fluctuation amplitude decreases and the frequency spectrum becomes more localized, with the $f$-$r$ spectrum concentrating around $\sim10~\mathrm{kHz}$ and the $\tilde{n}/n$ level becoming smaller (Fig. \ref{fig:Rm variation of cnoidal wave at 650G 2e-3 mbar}(f-g)). The $k_\theta$-$f$ spectrum (Fig. \ref{fig:Rm variation of cnoidal wave at 650G 2e-3 mbar}(h)) correspondingly exhibits fewer dominant coherent modes, while the bi-coherence spectrum shows weaker but still visible nonlinear coupling, indicating that mode-mode interaction persists but with reduced strength. It is also observed that strong nonlinear interaction is mainly confined up-to the $\sim6~\mathrm{kHz}$ frequency range (Fig. \ref{fig:Rm variation of cnoidal wave at 650G 2e-3 mbar}(i)), beyond which the coupling becomes significantly weaker, suggesting that the reduced gradients limit the generation of higher harmonic components. When $R_m$ is further decreased from $33$ to $17$, the background gradients become even weaker (Fig. \ref{fig:Mean profile at 650 G 2e-3 mbar}) (almost 50$\%$ of the initial amplitude). In this regime, the fluctuation spectra evolve toward a broadband nature, where most of the spectral power is distributed in the $12$ to $15~\mathrm{kHz}$ range (Fig. \ref{fig:Rm variation of cnoidal wave at 650G 2e-3 mbar}(k)). The $k_\theta$-$f$ spectrum shows a broader frequency distribution with very few distinct coherent modes (Fig. \ref{fig:Rm variation of cnoidal wave at 650G 2e-3 mbar}(m)), primarily around $13$-$15~\mathrm{kHz}$ and $\sim19.3~\mathrm{kHz}$, and the bi-coherence analysis indicates substantially diminished nonlinear coupling (Fig. \ref{fig:Rm variation of cnoidal wave at 650G 2e-3 mbar}(n)). Thus, the progressive reduction of density and velocity gradients ($\nabla n$ and $\nabla V_{E\times B}$; $R_m=50 > R_m=33 > R_m=17$) leads to a systematic decrease in fluctuation amplitude, spectral coherence, and nonlinear wave - wave interactions. Finally, the raw fluctuation signals shown in Fig.~\ref{fig:Mean profile at 650 G 2e-3 mbar}(g- i) clearly demonstrate that the waveform structure itself is strongly influenced by the strength of the background gradients. For sufficiently large gradients, the fluctuations exhibit quasi-periodic sawtooth-like structures that resemble nonlinear drift-acoustic modes and can be described by cnoidal-type solutions of the Korteweg - de Vries (KdV) equation. Such cnoidal waveforms arise when the nonlinear steepening of the drift-wave perturbation is balanced by dispersive effects, allowing the system to reach a coherent nonlinear state. The strong density and velocity gradients provide the free energy necessary to sustain this nonlinear balance, leading to the formation of periodic nonlinear structures with clear harmonic interactions. However, as the profile gradients decrease, this balance between nonlinearity and dispersion weakens. Consequently, the coherent cnoidal structures gradually disappear and the fluctuations evolve toward broadband turbulence with smaller amplitudes and weaker nonlinear coupling. Fig. \ref{fig: PSD and profile variation}a shows the variation of the density gradient and the $E\times B$ drift velocity for different plasma conditions obtained by changing the magnetic field and $R_m$. When the magnetic field is increased from $550~\mathrm{G}$ to $650~\mathrm{G}$ at $R_m=50$, both the density gradient and the $E\times B$ velocity increase significantly,  indicating a steepening of the background plasma profiles and stronger shear flow. In contrast, when $R_m$ is reduced from $50$ to $33$ and further to $17$ at $650~\mathrm{G}$, both the density gradient and the $E\times B$ velocity decrease systematically, implying a progressive reduction in the available free energy that drives drift-wave instabilities. The corresponding power spectral density (PSD) of the $\tilde{n}$ fluctuations is shown in Fig. \ref{fig: PSD and profile variation}(b-e). For the strong-  gradient case ($650~\mathrm{G}, R_m=50$), the spectrum exhibits a prominent fundamental peak around $\sim2~\mathrm{kHz}$ along with several harmonics, indicating the presence of coherent nonlinear modes. Such discrete spectral structures suggest strong nonlinear wave--wave interactions and are consistent with the formation of cnoidal-like nonlinear drift-wave structures. When the gradients are moderately reduced ($650~\mathrm{G}, R_m=33$), the fundamental peak remains visible but the harmonic components become weaker, suggesting a reduction in nonlinear coupling. For the weakest gradient case ($650~\mathrm{G}, R_m=17$), the spectrum becomes broadband with most of the power concentrated in the $10$--$15~\mathrm{kHz}$ band and no clear harmonic structure, indicating a transition toward a more turbulent fluctuation state. The reference case ($550~\mathrm{G}, R_m=50$) also shows coherent peaks but at a lower fundamental frequency ($\sim1~\mathrm{kHz}$), demonstrating that increasing the magnetic field shifts the characteristic fluctuation frequency. Overall, the figure highlights a clear connection between the background profile gradients and the nature of the fluctuations, where stronger gradients favor coherent nonlinear drift-wave structures while weaker gradients lead to broadband turbulence. This behavior indicates that a sufficiently strong background gradient is a necessary condition for sustaining the nonlinear state of drift waves and for the generation of cnoidal-type periodic waveforms in the plasma. In other words, the experimental observations demonstrate that the transition from coherent nonlinear drift-wave structures to broadband turbulence is closely controlled by the magnitude of the background profile gradients, which regulate the available free energy and the strength of nonlinear interactions in the system. 
\section{Physical Interpretation of the Observed Cnoidal-like Structures}
Drift waves are low-frequency electrostatic modes in magnetized plasmas driven by equilibrium density gradients. In the linear regime, these waves remain small in amplitude; however, as the instability grows, nonlinear effects become significant and modify the wave dynamics. The nonlinear evolution is governed by the interplay of vector nonlinearity due to $\mathbf{E}\times\mathbf{B}$ advection and scalar nonlinearity arising from the nonlinear Boltzmann response of electrons. While the vector nonlinearity leads to vortical structures and turbulence, the scalar nonlinearity introduces wave steepening, which, when balanced by dispersive effects associated with ion polarization drift, can result in coherent nonlinear structures such as solitons and cnoidal wave trains. In this context, analytical studies by Saleem \textit{et al.}~\cite{saleem2026exploring} have shown that, when the ion vorticity term is neglected, the Hasegawa--Mima equation~\cite{hasegawa1978pseudo} reduces to a nonlinear form that admits soliton solutions. Furthermore, under the assumption that the normalized electrostatic potential depends only on a single spatial coordinate along the predominant propagation direction, the equation further reduces to a one-dimensional KdV equation. Earlier theoretical works by Oraevskii \textit{et al.}~\cite{Oraevskii1969}, Petviashvili~\cite{Petviashvili1977}, and Lakhin \textit{et al.}~\cite{lakhin1988drift} demonstrated that finite-amplitude drift waves can evolve into regular, sawtooth-like structures resembling a train of solitons indicating a strongly nonlinear regime. Physically, the quadratic nonlinearity leads to wave steepening and the formation of sharp density gradients, while higher-order (cubic) nonlinearities become important at larger amplitudes, modifying the waveform and preventing wave breaking by redistributing energy within the structure; the balance of these nonlinear effects with dispersion gives rise to stable localized or periodic structures such as solitons and cnoidal waves. In the present experiment, the observed density fluctuations exhibit a finite amplitude [$\tilde{n}/n \sim (4$--$10)\%$] and are accompanied by a maximum mean $\mathbf{E}\times\mathbf{B}$ drift velocity of $\sim 3272~\mathrm{m/s}$ at the location of peak velocity shear. The fluctuation profile is characterized by pronounced sawtooth-like features with sharp peaks, along with a large normalized ratio [$\sim (4$--$8)$] between normalized density and potential fluctuations, clearly indicating a strongly nonlinear regime. Furthermore, both the experimentally measured density and potential fluctuation signals are well fitted by a cnoidal function with modulus $k \sim 0.85$--$0.99$, providing evidence for the presence of nonlinear periodic wave structures. As a result, the ion-acoustic mode nonlinearly interacts with drift-wave modes, leading to the formation of drift-acoustic cnoidal waves. These observations  qualitatively agree with the theoretical framework of drift soliton formation, suggesting that the measured structures correspond to nonlinear drift soliton or cnoidal-type coherent structures arising from the balance of quadratic and higher-order nonlinearities with dispersion.

\section{Discussion $\&$ conclusions}
\noindent
The present results demonstrate that the evolution of drift-wave fluctuations and the emergence of coherent structures are governed by the background plasma gradients, which control the available free energy and the relative strength of nonlinear interactions. By systematically varying the magnetic field and magnetic field ratio $R_m$, modifications in the density gradient and $\mathbf{E}\times\mathbf{B}$ velocity shear are achieved, leading to distinct changes in the fluctuation dynamics. At high profile gradients (higher magnetic field and larger $R_m$), the plasma exhibits enhanced density and electric field gradients, resulting in increased $E\times B$ flow shear and a corresponding rise in fluctuation amplitude. In this regime, the fluctuation spectra exhibit ion acoustic modes along with multiple higher harmonics, as well as drift wave modes in the $8$--$10~\mathrm{kHz}$ frequency range, and the bi-coherence analysis reveals strong nonlinear wave--wave coupling. The presence of discrete spectral peaks and harmonic interactions, along with wavelet signatures of temporally localized but coherent power, indicates the formation of nonlinear coherent structures. Consistently, the raw fluctuation signals shows quasi-periodic sawtooth-like waveforms, which are well described by cnoidal-type solutions, suggesting that the system attains a nonlinear state where wave steepening due to scalar nonlinearity is balanced by dispersive effects associated with ion polarization drift. As both the density gradient and velocity shear weaken, a systematic reduction in fluctuation amplitude and nonlinear coupling strength has been observed. In the intermediate regime, coherent modes persist but with reduced harmonic content and weaker bi-coherence levels, indicating decrease of non-linearity of the system. Further reduction of gradients leads to a transition toward broadband spectra, where the fluctuation power spreads over a wider frequency range and distinct coherent modes disappear. In this regime, the bi-coherence analysis shows feeble non linear coupling, suggesting that the nonlinear interactions are no longer sufficiently strong to sustain coherent structures. Correspondingly, the raw fluctuation signals lose their sawtooth-like character and evolve into irregular, turbulent waveforms. This transition is further supported by the estimate of the turbulent auto-decorrelation time of the $\tilde{n}$ signal at locations where nonlinear structures are observed. For the case of $550~\mathrm{G},~R_m=50$, the decorrelation time is found to be $\sim26~\mathrm{ms}$ with $k_\theta \sim 0.9~\mathrm{cm^{-1}}$ corresponding to the dominant low-frequency mode at $\sim1.1~\mathrm{kHz}$, indicating the presence of large-scale coherent structures. In contrast, for $650~\mathrm{G},~R_m=17$, where the profile gradients are significantly weaker, the decorrelation time decreases to $\sim0.01~\mathrm{ms}$ with $k_\theta \sim 1.8~\mathrm{cm^{-1}}$ for a higher dominant frequency of $\sim15~\mathrm{kHz}$. This reduction in decorrelation time, accompanied by an increase in wavenumber and frequency, clearly indicates a transition from large-scale coherent structures to small-scale turbulent fluctuations. These observations can be interpreted in terms of the competition between vector and scalar nonlinearities. For sufficiently strong gradients, the scalar nonlinearity associated with the nonlinear density response dominates, leading to wave steepening and the formation of periodic nonlinear wave trains. The experimentally measured ratio $V_{E\times B}/C_s$ is found to be $\sim1.1$ for $B_m=550~\mathrm{G},~R_m=50$, $\sim1.77$ for $B_m=650~\mathrm{G},~R_m=50$, $\sim1.48$ for $B_m=650~\mathrm{G},~R_m=33$, and $\sim1.69$ for $B_m=650~\mathrm{G},~R_m=17$ at $2\times 10^{-3}~\mathrm{mbar}$. These values follow the trend of variation in the background profile gradients shown in Fig.~\ref{fig: PSD and profile variation}a, indicating a close connection between flow dynamics and available free energy. The ratio $V_{E\times B}/C_s$ is of order unity suggests that the nonlinear advective velocity is comparable to the ion sound speed, placing the system in a regime where nonlinear steepening is significant while dispersive effects remain effective. This balance favors the formation of stable nonlinear periodic structures such as cnoidal waves. In contrast, as the profile gradients weaken, the nonlinear drive is reduced, leading to a breakdown of this balance and a transition from coherent structures to broadband turbulent fluctuations. Thus, the experimental results establish a clear link between background profile gradients and the nature of drift-wave dynamics: strong gradients favor the formation of coherent nonlinear structures such as cnoidal waves, while reduced gradients lead to a transition toward turbulent states. This transition reflects the weakening of the balance between nonlinearity and dispersion, which is essential for sustaining nonlinear drift soliton or cnoidal-type structures in magnetized plasmas.

\begin{acknowledgments}
\noindent
The authors would like to express their sincere gratitude to Kalpesh Doshi, Jignesh Patel, Kirti Mahajan, Manisha Bhandankar, Ritesh Sugandhi, Minsha Shah, Praveenlal, and Sandip Das from the Institute for Plasma Research (IPR), India, for their valuable experimental support. AS is grateful to the Indian National Science Academy (INSA) for the INSA Honorary Scientist position.
 
\end{acknowledgments}
\bibliographystyle{unsrt}
\bibliography{References}% Produces the bibliography via BibTeX.

\end{document}